\documentclass{pspum-l}

\usepackage{amsmath,amssymb,amsfonts}
\usepackage{graphicx}
\usepackage{caption}
\usepackage{subcaption}

\theoremstyle{definition}

\theoremstyle{remark}

\numberwithin{equation}{section}

\begin{document}

\title{Mirror Symmetry in Flavored Affine $D$-type Quivers}

\author{Anindya Dey}

\curraddr{Theory Group, University of Texas at Austin}

\email{anindya@physics.utexas.edu}

\thanks{The author would like to thank Jacques Distler for collaboration \cite{Dey:2013nf} on which much of this material is based. The author would also like to thank Peter Koroteev and Noppadol Mekareeya for some of the figures. This work (UTTG-01-14) is supported by the National Science Foundation under Grant Numbers PHY-1316033 and PHY-0969020. }

\date{January 15, 2014}


\begin{abstract}
We present non-trivial checks of three dimensional mirror symmetry for $\mathcal{N}=4$, $\hat{D}_N$ quiver gauge theories with unitary gauge groups using partition function on a round sphere. Type IIB (Hanany-Witten) realization of these theories and their mirror duals (as world volume gauge theories on coincident D3 branes) involve 1/2-BPS boundary conditions associated with orbifold and orientifold  5-planes respectively, in addition to NS5 and D5 branes. We demonstrate that partition function for a given quiver in this class may be decomposed into distinct contributions from the aforementioned Type IIB ingredients. As a byproduct of this computation, we find a convenient way of deriving the mirror map for a given pair of dual theories.
\end{abstract}

\maketitle

\section{Introduction}
Mirror symmetry in $\mathcal{N}=4$ supersymmetric affine quiver gauge theories in three dimensions have been extensively studied in the literature \cite{Intriligator:1996ex, deBoer:1996mp, Hanany:1996ie}. In the Type IIB picture, these theories arise as world volume gauge theories on coincident D3 branes wrapping $\mathbb{R}^{2,1} \times L$ -- $L$ being a compact direction -- with certain special $1/2$-BPS  boundary conditions at the two ends. In Type IIA, these can be obtained as world volume theories on coincident D2 branes with parallel D6 branes (and possibly a O6 plane) wrapping a transverse ALE (Asymptotically Locally Euclidean) space. The M-theory lift of this Type IIA scenario therefore involves coincident M2 branes probing a product of ALE spaces in the transverse direction with appropriate G-fluxes. As is evident from the Type IIA description, Higgs branches of such theories arise as moduli spaces of instantons for classical gauge groups on ALE spaces \cite{Dey:2013fea}.

Aside from other interesting features, such theories provide a rich laboratory for studying dualities in supersymmetric QFTs. Mirror symmetry, a  particularly important duality for $\mathcal{N} = 4$ theories in three dimensions, acts by simply exchanging the Coulomb and the Higgs branches of a pair of theories. It is a classic example of a IR duality which involves two or more theories with completely different UV description flowing to the same superconformal point in the IR. 

Recent progress in  localization techniques \cite{Pestun:2007rz, Kapustin:2009kz} has provided an unprecedented opportunity to study such dualities by  computing RG-independent supersymmetric observables  $\emph{exactly}$ on both sides of the duality and checking that the results agree. In particular, partition function on a round sphere  turns out to be an extremely effective tool for studying dualities in three dimensions and was used in \cite{Kapustin:2010xq} to provide a non-trivial check for mirror symmetry in affine $A$-type ($\hat{A}_N$) quivers. In addition, it was shown that partition function of any quiver of this type admits a convenient decomposition in terms of contributions from NS5 and D5 branes. Furthermore, these contributions are precisely exchanged under S-duality implemented at the level of the partition function, as one should expect.

In the present work, we will demonstrate how mirror symmetry in flavored affine $D$-type ($\hat{D}_N$) quiver gauge theories can be analyzed using partition functions  on a round sphere. Aside from providing extremely non-trivial checks on mirror symmetry for this class of quivers, we show that the partition function may be decomposed again into appropriate ``building blocks" which are in one-to-one correspondence with the ingredients of the Type IIB set-up.  In addition to contributions of NS5 and D5 branes familiar from the $\hat{A}_N$ case, these include contributions from certain 1/2-BPS boundary conditions involving orbifold/orientifold 5-planes which we shall discuss in detail in the paper.

\section{Rudiments of Mirror Symmetry and $S^3$ Partition Function }
In this section, we summarize the basic field theory ingredients necessary for our story. We will also briefly discuss $S^3$ partition functions for $\mathcal{N}=4$ gauge theories obtained using localization techniques.

\subsection{Field Theory Fundamentals.}
$\mathcal{N}=4$ supersymmetry in $D=3$ has 8 real supercharges, which are doublets of $Spin(2,1)\sim SL(2,\mathbb{R})$ and transform as $(2,2)$ under the R-symmetry group, $SU(2)_R \times SU(2)_L$. Under the R-symmetry group $SU(2)_L \times SU(2)_R$, the relevant supermultiplets transform in the following way:\\
Vector multiplet:$(3\oplus1,1)+\mathbf{(2,2)}$\\
Half-hyper multiplet:$(1,2)+\mathbf{(2,1)}$

The half-hypers transform in pseudo-real representations of the gauge group. A $\mathcal{N}=4$ hypermultiplet in three dimensions consists of two copies of half-hypers.

For the 3D vector multiplets, the bosons consist of an $SU(2)_L$-triplet of scalars and a gauge boson, which is an R-symmetry singlet. Strictly speaking, the latter can be dualized to a circle-valued scalar only for an abelian gauge field. In that case, we obtain 4 scalars,transforming overall as $(3\oplus1,1)$. Nevertheless, for a non-abelian gauge field, the gauge symmetry (on either the Coulomb or Higgs branches) is higgsed to $\emph{at most}$ an abelian subgroup. So it is useful to carry over this counting in describing the low-energy theory.

Since the matter content in these supermultiplets is not symmetric with respect to the exchange $SU(2)_R \leftrightarrow SU(2)_L$, there can be ``twisted'' multiplets where $SU(2)_R$ and $SU(2)_L$ are exchanged.

\subsection{IR Behavior and Mirror Symmetry.} We will refer to a pair of mirror dual theories as the A-model and the B-model \footnote{This nomenclature, which dates back to \cite{deBoer:1996mp}, has nothing to do with the 2D homological mirror symmetry.}.
Mirror symmetry in three dimensions for $\mathcal{N}=4$ theories has the following broad features:
\begin{itemize}
\item The duality exchanges the Coulomb and the Higgs branch of the A and the B-model:\\
\begin{equation}
\begin{split}
&\mathcal{M}^{(A)}_{Coulomb}= \mathcal{M}^{(B)}_{Higgs}\\
&\mathcal{M}^{(A)}_{Higgs}= \mathcal{M}^{(B)}_{Coulomb}\\
\end{split}
\end{equation}
Note that it is a strictly IR duality -- two theories with very different UV descriptions  flowing to the same $\mathcal{N}=4$ SCFT in the IR.

\item Mirror Symmetry exchanges $SU(2)_R$ and $SU(2)_L$ -- thereby exchanging background vector and twisted vector multiplets. This naturally implies exchange of hypermultiplet masses and FI parameters under the duality.

\item The precise transformation between masses on one side of the duality and FI parameters on the other is captured by a linear map -- often referred to as the ``Mirror Map". There's also a map between the chiral operators on the Coulomb branch of the A-model and the Higgs branch of the B-model as expected.
\item Mirror Symmetry is a direct consequence of S-duality in the Type IIB setting. Under S-duality, NS5 and D5 branes are exchanged while D3 branes are self-dual. Similarly, an orbifold 5-plane is S-dual to a O$5^0$ planes (i.e. O$5^{-}$ plane coincident with a D5) -- this plays a crucial role in mirror symmetry for affine $D$-type quivers.

\end{itemize}

\subsection{$S^3$ partition function} Let us briefly summarize the rules for writing down the $S^3$ partition function of any given $\mathcal{N}=4$ quiver gauge theory. 
\begin{itemize}
\item Localization ensures that the partition function of the theory reduces to a matrix integral  in terms of the real adjoint scalar $s$ of the  3d $\mathcal{N}=2$ vector multiplet . Any such partition function may be written as 
\begin{equation}
Z=\int  \frac{d^k s}{|\mathcal{W}|} \prod_{\alpha} \alpha(s) \exp{S_{\text{cl}}[s]}\, Z_{\text{1-loop}}[s]\,,
\end{equation}
Note that one can make $s$ lie in the Cartan subalgebra of the gauge group using a constant gauge transformation.  In the above formula $\prod_{\alpha} \alpha(s)$  is the product  over all roots of the gauge group with $s$ being in the Cartan of the gauge group - this is simply the Vandermonte determinant. Finally, one needs to divide by the order of the Weyl group $|\mathcal{W}|$  to account for the residual gauge symmetry.

\item Consider first the contribution of vector multiplets in the $\mathcal{N} = 4$ theory to the classical and 1-loop part of the partition function. The classical contribution for every 
$U(1)$ factor in the gauge group is 
\begin{equation}
S^{\text{FI}}_{\text{cl}}= 2\pi i \eta\, \text{Tr}(s)\,,
\end{equation}
where $\eta$ is a FI parameter. To the 1-loop part, every $\mathcal{N}=4$ vector multiplet contributes
\begin{equation}
Z^v_{\text{1-loop}}=\prod_{\alpha} \frac{\sinh^2{\pi \alpha(s)}}{\pi \alpha(s)}\,,
\end{equation}
where the product is over all the roots of the Lie algebra of G. 

\item The contribution of each $\mathcal{N}=4$ hypermultiplet is
\begin{equation}
Z^h_{\text{1-loop}}=\prod_{\rho} \frac{1}{\cosh{\pi \rho(s+m)}}\,,
\end{equation}
where the product extends over all the weights of the representation R of the gauge group G and $m$ is a real mass parameter. 

A factor of Vandermonte determinant appears in the measure as a result of gauge fixing $s$. This exactly cancels with the denominator of the 1-loop contribution of the vector multiplet for each factor in the gauge group and can therefore be ignored  in the matrix integral.

Given a quiver gauge theory, one can simply read off the independent mass deformations from the partition function -- these correspond to the masses that cannot be removed by constant shifts in the integration variable $s$.

\end{itemize}

\section{Mirror Symmetry and $S^3$ partition function} Let us state the basic idea underlying our computation. Given a pair of conjectured dual theories, one may compute partition functions of such theories as functions of masses and FI parameters and check whether they agree up to some overall field-independent phase. Therefore, the strategy is to start with the partition function of the A-model and implement some clever change of variables to show that it is equivalent to the partition function of the B-model. Note that this change of variables essentially corresponds to implementing S-duality at the level of the partition function. Comparison of partition functions of the conjectured dual theories as functions of masses and FI parameters naturally allows one to read off the ``Mirror Map" -- the linear map between masses (FI parameters) on one side and FI parameters (masses) on the other.

For generic rank of the gauge group at every node in a given quiver, this manipulation is difficult and requires some very non-trivial identities involving hyperbolic functions, which we will discuss for the case of  $\hat{A}_N$ and  $\hat{D}_N$ quivers.

\subsection{$\hat{A}_N$ Quivers: NS5 and D5 Contributions }
\begin{figure}[htbp]
\begin{center} 
\includegraphics[height=2.0in]{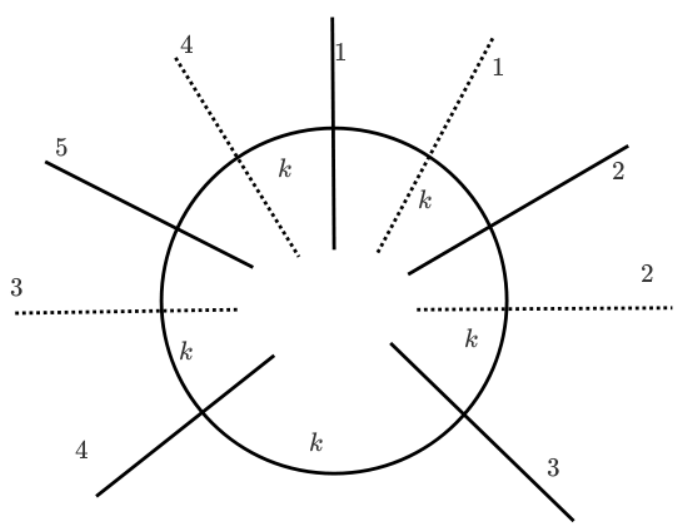}
\caption{Type IIB set-up for a $\hat{A}_5$ quiver with gauge group $U(k)^5$ and fundamental hypers distributed as $\{1,1,0,1,1\}$}-- the circle denotes $k$ coincident D3 branes wrapped along $S^1$, solid lines denote NS5 while dotted lines denote D5 branes.
\label{fig0al}
\end{center}
\end{figure}

Affine $A$-type quivers arise in Type IIB string theory as world volume gauge theories on D3 branes wrapping a $S^1$ with NS5 and D5 branes inserted at different positions on this circle direction, see figure \ref{fig0al} for an example. There are two crucial identities needed to manipulate partition functions of this class of quiver gauge theories, namely:
\begin{itemize}
\item \textbf{Fourier Transform of Hyperbolic Secant}
\begin{equation}
\int \frac{e^{2\pi i x z}}{\cosh{\pi z}} dx = \frac{1}{\cosh{\pi x}} \label{ftcosh}
\end{equation}

\item \textbf{Cauchy's Determinant Identity} (see \cite{Dey:2013nf, Kapustin:2010xq} for details)
\begin{equation}
\sum _{\rho} (-1)^{\rho} \frac{1}{\prod_{i}\cosh{(x_i -y_{\rho(i)})}}=\frac{\prod_{i <j} \sinh{(x_i-x_j)} \sinh{(y_i -y_j)}}{\prod_{i,j}\cosh{(x_i -y_j)}} \label{detid}
\end{equation}

\end{itemize}

The partition function for a $\hat{A}_N$ quiver gauge theory with gauge group $U(k)^N$ and $L$ fundamental hypers distributed in some arbitrary fashion among the $N$ gauge groups can be schematically written as
\begin{equation}
Z_{\hat{A}_N}=\int  \prod^{N+L}_{\alpha=1} d^k s_{\alpha} \mathcal{Z}_{NS5} (s_1, s_2).... \mathcal{Z}_{D5}(s_k, s_{k+1})...\mathcal{Z}_{NS5}(s_{N+L}, s_{1}).
\end{equation}
where the functions $\mathcal{Z}_{NS5}$ and $\mathcal{Z}_{D5}$ are contributions of individual NS5 and D5 branes to the partition functions. Explicitly, $\mathcal{Z}_{NS5}$ is given as 
\begin{equation}
\begin{split}
\mathcal{Z}_{NS5}(s_M,s_{M+1})&=\frac{\prod^k_{i <j} \sinh{\pi (s^i_M-s^j_{M})} \sinh{\pi(s^i_{M+1} -s^j_{M+1})}}{\prod^k_{i,j=1}\cosh{\pi(s^i_{M} -s^j_{M+1})}}\\
&=\int d^k\tau_M \sum _{\rho} (-1)^{\rho} \prod^k_{i=1} \frac{e^{2\pi i \tau^i_M (s^i_M -s^{\rho(i)}_{M+1})}}{\cosh{\pi \tau^i_M}} \label{NS5}
\end{split}
\end{equation}
where in the final step we have used equation (\ref{ftcosh}) and equation (\ref{detid}). The function $\mathcal{Z}_{D5}$ reads
\begin{equation}
\begin{split}
\mathcal{Z}_{D5}(s_M,s_{M+1})&=\sum _{\rho} (-1)^{\rho} \prod^k_{i=1} \frac{\delta(s^i_M-s^{\rho(i)}_{M+1})}{\cosh{\pi s^i_M}}\\
&=\int d^k\tau_M  \sum _{\rho} (-1)^{\rho} \prod^k_{i=1} \frac{e^{2\pi i \tau^i_M(s^i_M-s^{\rho(i)}_{M+1})}}{\cosh{\pi s^i_M}} \label{D5}
\end{split}
\end{equation}
Note how in both cases we have introduced some auxiliary variables $\tau^i$ via Fourier transformation. Implementing S-duality at the partition function level now simply amounts to integrating over the variables $s^i$ and writing the partition function in terms of $\emph{only}$ the auxiliary variables $\tau^i$. One can easily check that this indeed exchanges the NS5 and D5 contributions in a given partition function as one would expect.  Written in terms of the auxiliary variables, the partition function can be readily identified as the correct partition function for the dual gauge theory with variables $\tau^i$ identified as vevs of real adjoint scalars in appropriate vector multiplets for the dual theory.

\subsection{$\hat{D}_N$ Quivers: Orbifold/Orientifold Contributions} 
\begin{figure}[htbp]
\begin{center}
\includegraphics[width=0.9\textwidth]{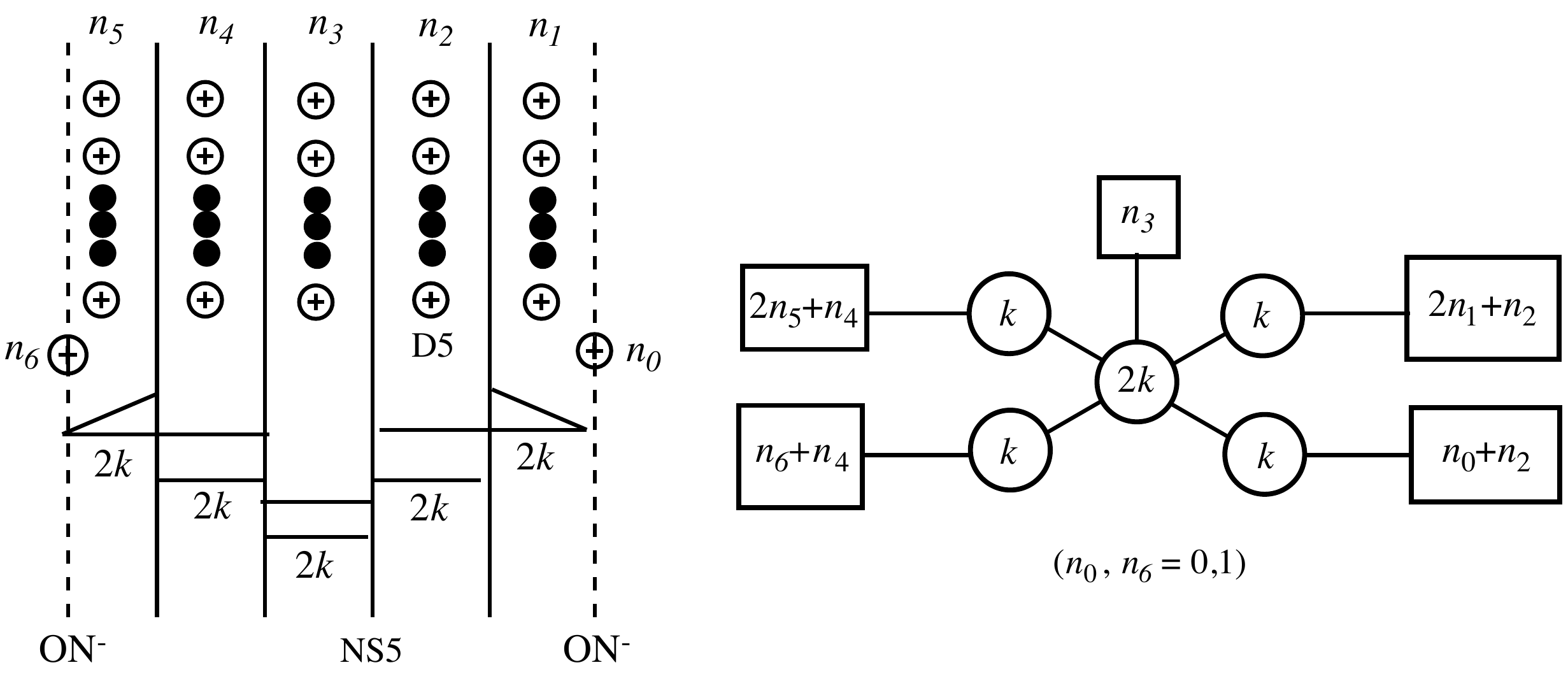}
\caption{Type IIB set-up of a generic flavored $\hat{D}_{4}$ quiver. The numbers $n_1, \ldots, n_5$ label the numbers of D5-branes at each interval, and the numbers $n_0, n_6=0,1$ labels the numbers of D5-branes stuck on each ON$^-$ plane. ON$^-$ plane and its adjacent NS5 coincide to give an orbifold 5-plane.}
\label{fig:BraneflavoredD4}
\end{center}
\end{figure}

In addition to  D5, NS5 and D3 branes, Type IIB description for affine $D$-type quivers involve orbifold 5-planes. An orbifold 5-plane is often understood as the combination of a NS5 and a coincident ON$^{-}$ plane (S-dual to a O5$^{-}$ plane) -- we refer the reader to \cite{Hanany:1999sj} for further details on this construction. The orbifold planes are oriented parallel to the NS5 branes. The compact direction wrapped by D3 branes is now a line segment with orbifold planes (with or without stuck D5 branes) at the two boundaries -- see figure \ref{fig:BraneflavoredD4} for an example. The mirror dual of an affine $D$-type quiver therefore arise from a Type IIB set-up where D3 branes wrap a line segment with O5$^0$ planes (O5$^-$ plane with a coincident D5) at the two boundaries with possibly some NS5 branes stuck to the O5$^-$ plane. NS5 and D5 branes are exchanged as usual under S-duality. Aside from equations (\ref{ftcosh}) and (\ref{detid}), there are two crucial identities needed to manipulate partition functions for this class of quiver gauge theories, namely:
\begin{itemize}
\item \textbf{Fourier Transform of Hyperbolic CoSecant}
\begin{equation}
\int \frac{e^{2\pi i x z}}{\sinh{\pi z}} dx = i\tanh{\pi x} \label{ftsinh}
\end{equation}
\item \textbf{A Variant of Cauchy's Determinant Identity}
\begin{equation}
\sum _{\rho} (-1)^{\rho} \frac{1}{\prod_{i}\sinh{(x_i -y_{\rho(i)})}}=\frac{\prod_{i <j} \sinh{(x_i-x_j)} \sinh{(y_i -y_j)}}{\prod_{i,j}\sinh{(x_i -y_j)}} \label{detidgen}
\end{equation}

\item \textbf{Schur's Pfaffian Identity} (see \cite{Dey:2013nf,Okada:2004Pf} for more details)
\begin{equation}
 Pf\left[\frac{\sinh{(x_p - x_l)}}{\cosh{(x_p + x_l)}} \right] =\left(\prod_{p<l} \frac{\sinh{(x_p - x_l)}}{\cosh{(x_p + x_l)}}\right) \label{pfid}
\end{equation}

\end{itemize}

The partition function of any flavored $\hat{D}_N$ quiver with unitary gauge groups can be schematically written as 
\begin{equation}
Z_{\hat{D}_N}=\int  \prod^{N+L-3}_{\alpha=1} d^k s_{\alpha} \mathcal{Z}^{(1)}_{bdry}(s_1)\mathcal{Z}_{NS5} (s_1, s_2)\ldots  \mathcal{Z}_{NS5}(s_{N+L-4}, s_{N+L-3})\mathcal{Z}^{(2)}_{bdry}(s_{N+L-3}).
\end{equation}
where $L$ is the number of fundamental hypers on the internal nodes of the quiver (i.e. nodes with Dynkin label 2) -- the dots indicate contributions from other NS5 and D5 branes as discussed for $\hat{A}_{N}$ quivers. $\mathcal{Z}^{(1)}_{bdry},\mathcal{Z}^{(2)}_{bdry}$ represent the contribution of the two boundaries to the partition function. 

\begin{figure}[h]
\begin{center}
\includegraphics[scale=0.3]{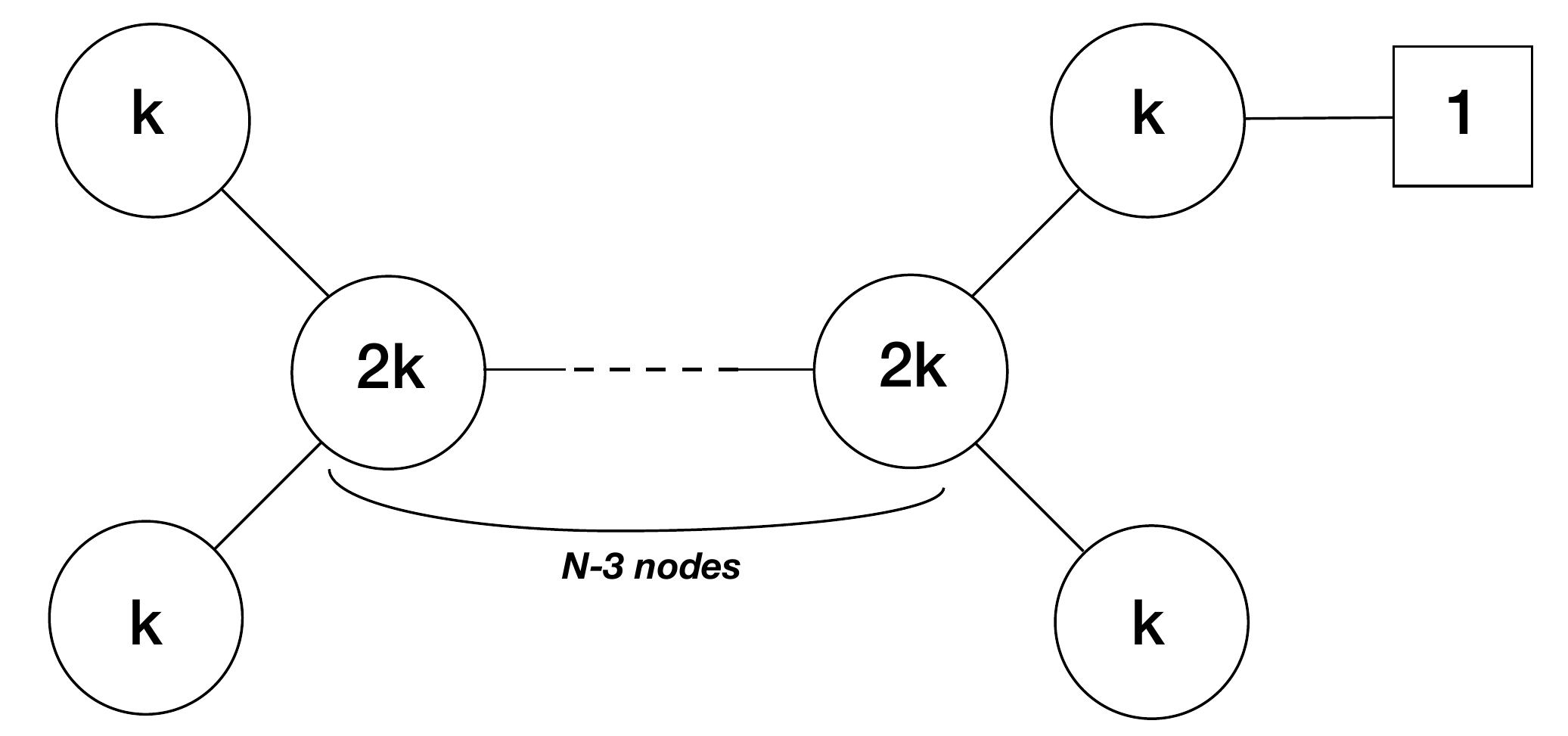} \quad \quad\quad \includegraphics[scale=0.5]{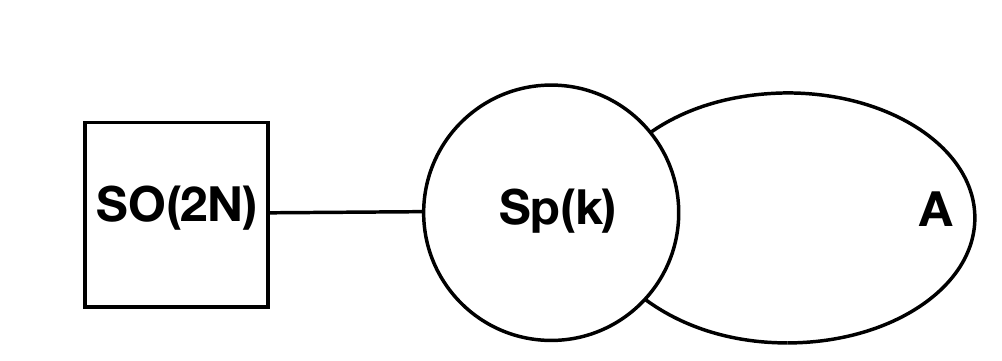}
\caption{$\hat{D}_N$ quiver with a single hyper on a boundary node. The mirror dual is a $Sp(k)$ gauge theory with $N$ flavors and one antisymmetric hyper. For $k=1$, the antisymmetric hyper is just a singlet.}
\label{figD1}
\end{center}
\end{figure}

\begin{figure}[h]
\begin{center}
\includegraphics[scale=0.3]{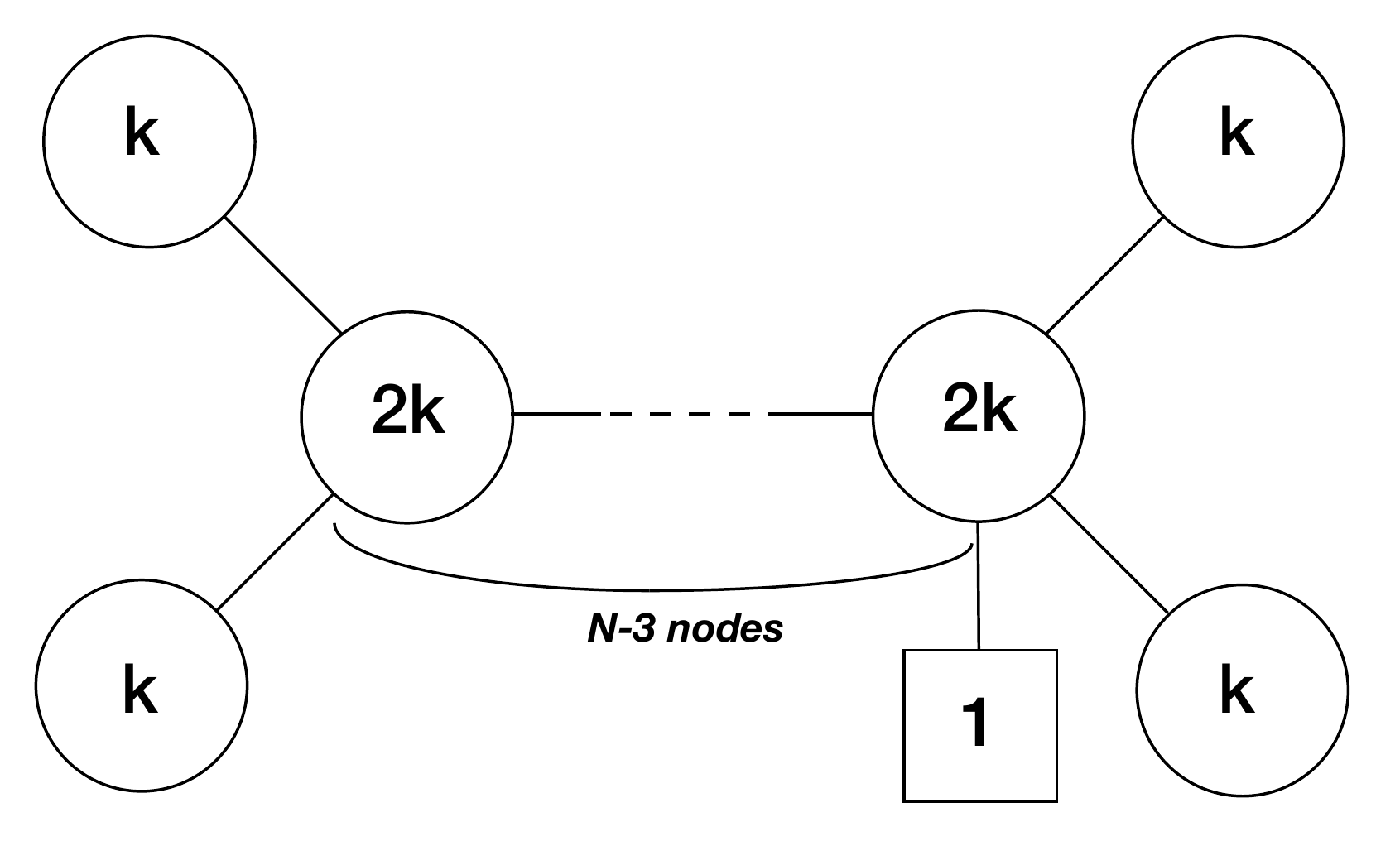} \quad\quad\quad \includegraphics[scale=0.5]{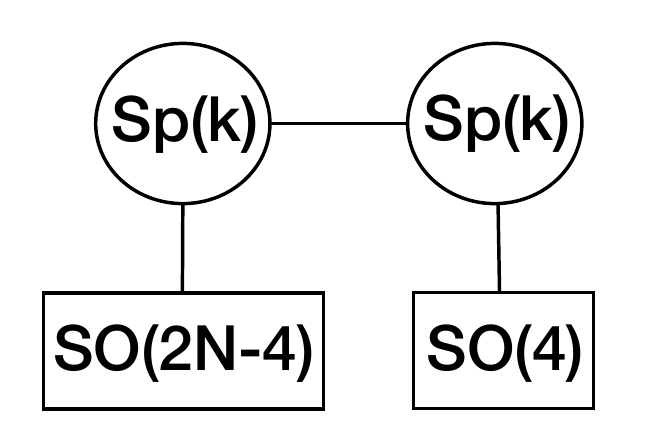}
\caption{$\hat{D}_N$ quiver with a single hyper on a boundary node. }
\label{figDin1}
\end{center}
\end{figure}

\begin{figure}[h]
\begin{center}
\includegraphics[scale=0.3]{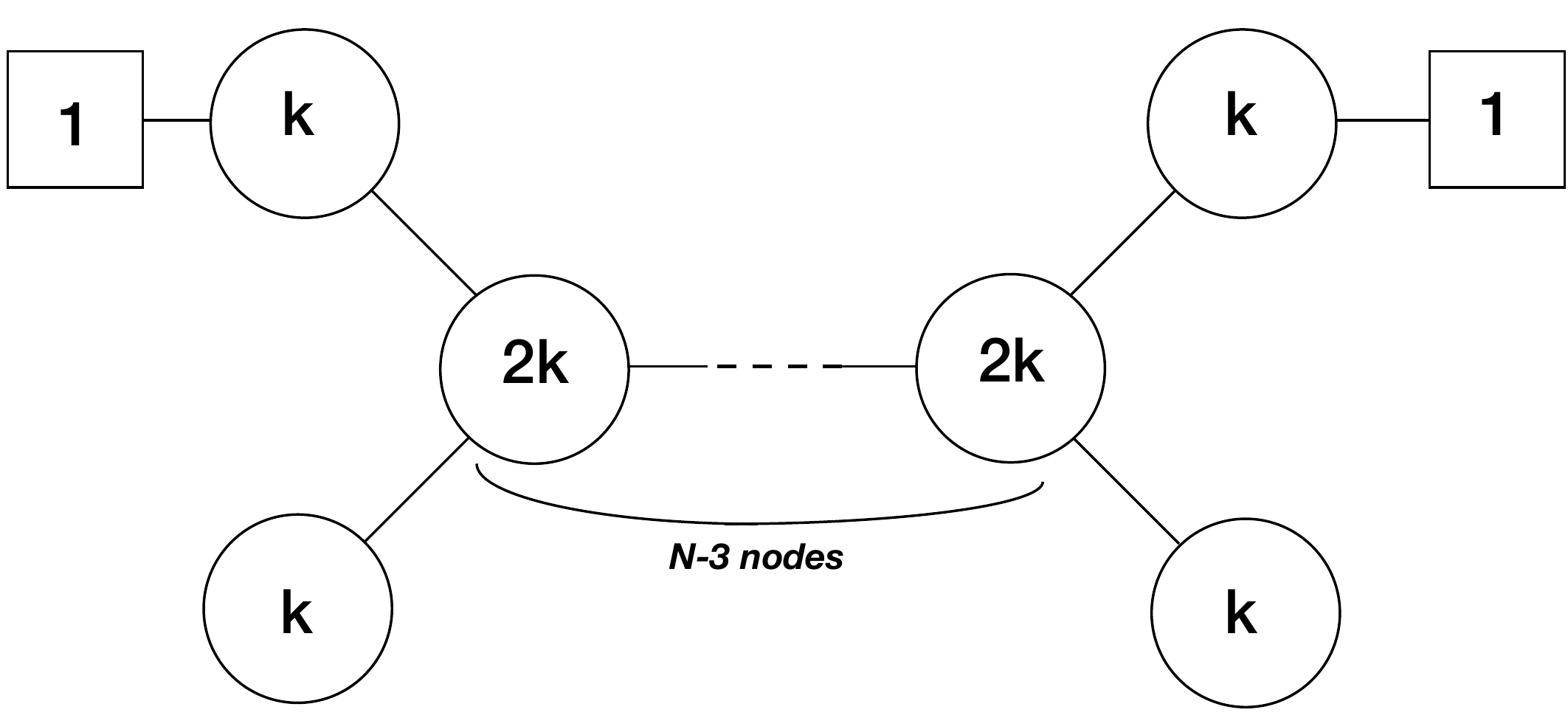}  \quad\quad\quad  \includegraphics[scale=0.5]{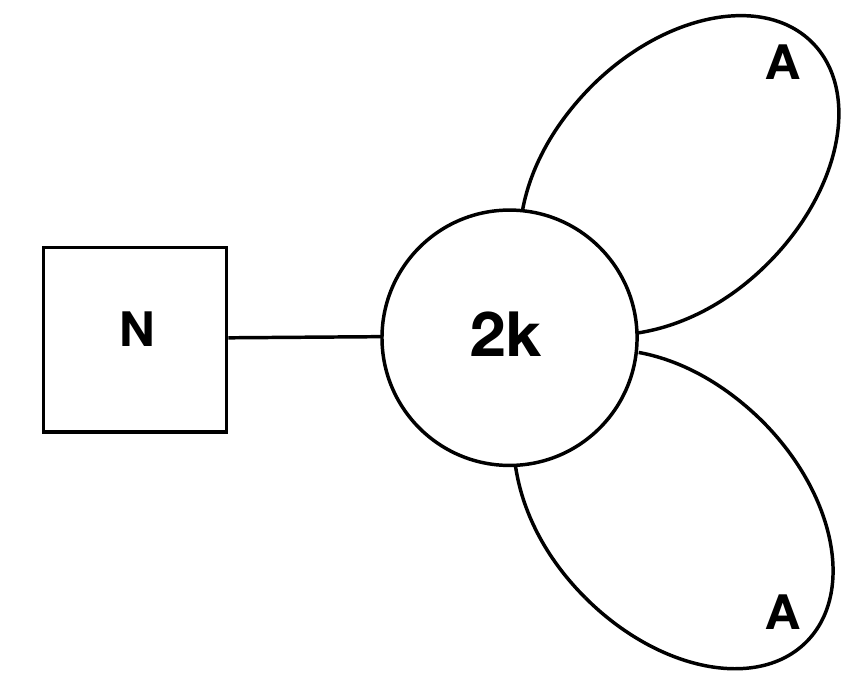} 
\caption{$\hat{D}_N$ quiver with two hypers on a boundary node. In this case, one has an orbifold 5-plane with a stuck D5 at each boundary.}
\label{figD2}
\end{center}
\end{figure}

\begin{figure}[h]
\begin{center}
 \includegraphics[scale=0.3]{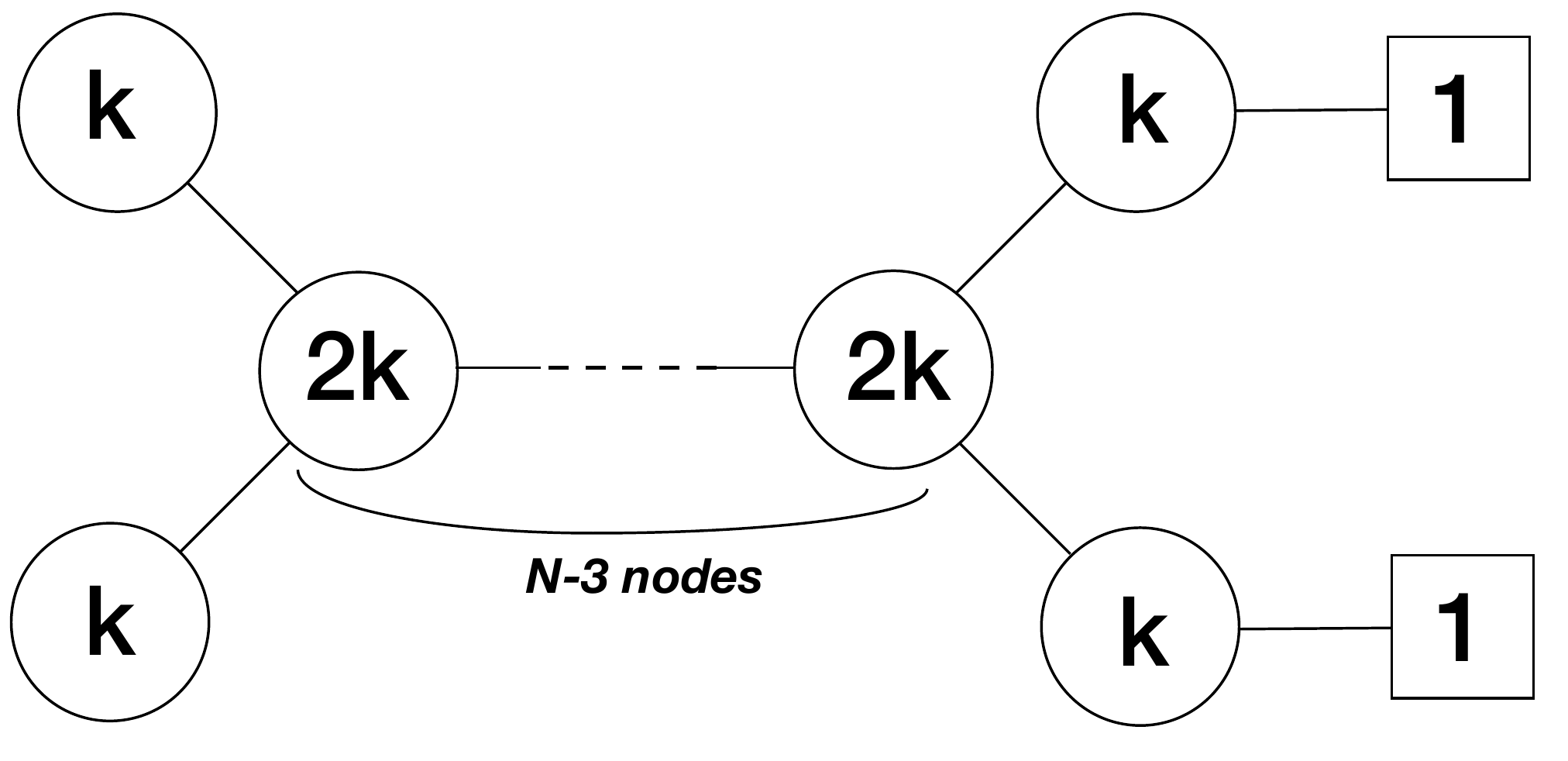} \quad\quad\quad \includegraphics[scale=0.5]{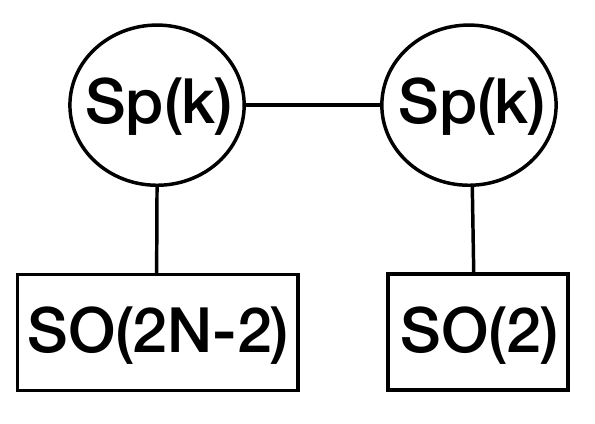}
\caption{$\hat{D}_N$ quiver with two hypers on a boundary node. In this case, one has an orbifold 5-plane without stuck D5 at each boundary. In addition, there is a single D5 brane between one of the orbifold planes and the nearest NS5 brane.}
\label{figD3}
\end{center}
\end{figure}

In the present discussion, we will restrict ourselves to the simpler cases where $n_1,n_5=0$ (quivers for which $n_1,n_5 \neq 0$ are discussed in \cite{Dey:2013nf}). For this subclass of quivers, one can have two possible boundary conditions:
\begin{itemize}
\item \textbf{Orbifold plane with no stuck D5 brane ($n_0=0$)}
\begin{equation}
\begin{split}
\mathcal{Z}_{Orb5}(s)&=\frac{\prod^k_{i <j} \sinh{\pi (s^i-s^j)} \sinh{\pi(s^{k+i} -s^{k+j})}}{\prod^k_{i,j=1}\sinh{\pi(s^i -s^{k+j})}}\\
&=i^k\int d^k\tau \sum _{\rho} (-1)^{\rho} \prod^k_{i=1} \frac{e^{2\pi i \tau^i (s^i -s^{k+\rho(i)})}}{\coth{\pi \tau^i}} \label{w/oD5}
\end{split}
\end{equation}

\item \textbf{Orbifold plane with a stuck D5 brane ($n_0=1$)}
\begin{equation}
\widetilde{\mathcal{Z}}_{Orb5}(s)=\mathcal{Z}_{Orb5}(s) \times \int d^k\tau \prod^k_{i=1}\frac{e^{2\pi i \tau^i s^i}}{\cosh{\pi \tau^i}} \label{wD5}
\end{equation}
\end{itemize}

where $\rho$ denotes permutation of the integers $\{1,2,\ldots,k\}$. Equations (\ref{w/oD5}) and (\ref{wD5}) along with contributions of the NS5 and D5 branes, as given in (\ref{NS5}) and (\ref{D5}) respectively, form all the ``building blocks'' necessary to construct the partition function of a flavored $\hat{D}_N$ quiver.
As discussed in the case of $\hat{A}_N$ quivers, S-duality is implemented by simply integrating over the variables $s^i$ and writing the partition function in terms of the auxiliary variables $\tau^i$ introduced in the above equations via Fourier transformation. This directly leads to the $S^3$ partition function of the dual gauge theory, as for example, in the infinite families of mirror pairs shown in figures \ref{figD1}-\ref{figD3}, thereby giving a non-trivial check on the conjectured duality. This procedure also allows one to read off the contribution of boundary conditions S-dual to those presented in equations (\ref{w/oD5}) and (\ref{wD5}) at the level of the $S^3$ partition function, i.e. O5$^0$ planes without and with stuck NS5 branes respectively. 
\begin{itemize}
\item \textbf{ O5$^0$ planes with no stuck NS5 brane}
\begin{equation}
\begin{split}
\mathcal{Z}_{O5^0}(s)=\frac{\prod^k_{i=1} \sinh{2 \pi s^i}\; \delta(s^i+s^{k+i})}{\prod^{2k}_{p=1} \cosh{\pi s^p}} \label{w/oNS5}
\end{split}
\end{equation}
\item \textbf{ O5$^0$ planes with stuck NS5 brane}
\begin{equation}
\begin{split}
\widetilde{\mathcal{Z}}_{O5^0}(s)&=C \sum_{\rho} (-1)^{\rho} \prod_i \frac{\tanh{\pi s^{\rho(k+i)}}}{\cosh{\pi(s^{\rho(k+i)}+s^{\rho(i)})}}\\
&=\frac{C}{\prod_p \cosh{\pi s^p}} \times  Pf\left[\frac{\sinh{\pi (s^p - s^l)}}{\cosh{\pi(s^p + s^l)}} \right] \\
&=\frac{1}{\prod_p \cosh{\pi s^p}} \times \prod_{p<l} \frac{\sinh{\pi (s^p - s^l)}}{\cosh{\pi(s^p + s^l)}}  \label{wNS5}
\end{split}
\end{equation}
\end{itemize}
where $C$ is a combinatorial factor and $\rho$ denotes permutation of the integers $\{1,2,\ldots,2k\}$.\\

Consider the mirror dual of the simplest possible flavored $\hat{D}_N$ quiver (in fact an infinite family of such quivers labeled by $k$ and $N$), i.e. with a single fundamental hyper as shown in figure \ref{figD1}. 

Type IIB set-up for this quiver involves an orbifold 5-plane at one boundary with a stuck D5 at one boundary and an orbifold 5-plane without any stuck D5 at the other boundary. One can therefore use the formulae for the appropriate boundary conditions as stated in equations (\ref{wD5}) and (\ref{w/oD5}), S-dualize the partition function and arrive at the mirror in figure\ref{figD1}. One may equivalently start from the mirror theory and using contributions of the appropriate boundary conditions (O5$^0$ planes with and without NS5) as given in equations (\ref{w/oNS5}) and (\ref{wNS5}) arrive at the particular $\hat{D}_N$ quiver. An important byproduct of this computation is to provide a convenient way of determining the mirror map which we present explicitly for this mirror pair.
\begin{equation}
\begin{split}
& m_1= (\eta_1-\eta_2), \\
&m_2 = (\eta_2+\eta_1),\\
&m_{\beta+2}=(\eta_2+\eta_1)+2\sum^{\beta}_{\alpha=1} \tilde{\eta}_{\alpha},\; \;\beta=1,2,..,N-3\\ 
&m_{N}=(\eta_2+\eta_1) +2\eta_4+2\sum^{N-3}_{\alpha=1} \tilde{\eta}_{\alpha}\\
&M_{AS} =\eta_1+\eta_2 +\eta_3 +\eta_4 + 2(\tilde{\eta}_1+....+\tilde{\eta}_{N-3})
\end{split}
\end{equation}
The above equation relates the masses of the B-model with the FI parameters of the A-model --- $\eta_i$ and $\tilde{\eta}_{\alpha}$ respectively denote the FI parameters associated with nodes of Dynkin label one and two respectively in the $\hat{D}_N$ quiver.

One can similarly treat a $\hat{D}_N$ quiver with a single fundamental hyper on one of the internal (Dynkin label 2) node. In this case, one has an orbifold 5-plane without any stuck D5 at each boundary. The aforementioned prescription for writing the partition function and S-dualizing leads to the mirror theory shown in figure \ref{figDin1}.

One may now engineer situations involving various combinations of these boundary conditions to obtain mirror pairs like the ones in figures \ref{figD2} and \ref{figD3}. The details of the partition function computation and determination of the mirror map in each case can be found in \cite{Dey:2013nf}. More complicated quivers with arbitrary distributions of fundamental hypers may be dealt with by simply putting together the basic ``building blocks" discussed above. For a discussion on some of these quivers, the reader is referred to section 6 and 7 of \cite{Dey:2013nf}.

\section{Conclusion} In this work, we have performed non-trivial checks on three dimensional mirror symmetry for $\mathcal{N}=4$, $\hat{D}_N$ quivers with unitary gauge groups using partition function on a round sphere. Comparing partition functions of the dual theories as functions of masses and FI parameters leads to a particularly convenient way of reading off the mirror map. We have shown that the partition function of a flavored $\hat{D}_N$ theory can be decomposed into contributions from NS5 branes, D5 branes and orbifold 5-planes (with or without stuck D5 branes) precisely as the Type IIB construction of such theories suggests. A simple prescription for S-dualization at the level of the partition function then leads to the partition function of the appropriate dual theory. This procedure naturally allows one to determine the contributions of orientifold 5-planes (with or without stuck NS5 branes), which appear in the dual of a flavored $\hat{D}_N$ theory, to the partition function. It would be interesting to understand how our story extends to mirror pairs whose Type IIB description involve O3 planes -- these, for example,  include $\hat{D}_N$ quivers with orthosymplectic gauge groups. A formulation that can take into account general 1/2-BPS boundary conditions at the partition function level in a manner discussed above may also be useful.

\bibliographystyle{amsalpha}

\end{document}